 \documentclass{elsart}

\usepackage{cite}

 \usepackage{graphicx}

\usepackage{amssymb}
\begin{document}
\journal{Physics Letters A}
\begin{frontmatter}


\title{Are generalized synchronization and noise--induced
synchronization identical types of synchronous behavior of chaotic
oscillators?}

\author[SSU]{Alexander~E.~Hramov},
\ead{aeh@cas.ssu.runnet.ru}
\author[SSU]{Alexey~A.~Koronovskii},
\ead{alkor@cas.ssu.runnet.ru}
\author[SSU]{Olga~I.~Moskalenko}
\ead{moskalenko@nonlin.sgu.ru}
\address[SSU]{Faculty of Nonlinear Processes, Saratov State %
University, Astrakhanskaya, 83, Saratov, 410012, Russia}

\begin{abstract}
This paper deals with two types of synchronous behavior of chaotic
oscillators --- generalized synchronization and noise--induced
synchronization. It has been shown that both these types of
synchronization are caused by similar mechanisms and should be
considered as the same type of the chaotic oscillator behavior.
The mechanisms resulting in the generalized synchronization are
mostly similar to ones taking place in the case of the
noise-induced synchronization with biased noise.
\end{abstract}

\begin{keyword}
coupled oscillators \sep chaotic synchronization \sep generalized
synchronization \sep noise--induced synchronization
\PACS 05.45.-a \sep 05.45.Xt \sep 05.45.Tp
\end{keyword}

\end{frontmatter}

Synchronization of chaotic oscillators has been intensively
investigated recently. The chaotic synchronization plays an
important role for the analysis of physiological and medicine
data, for a chaotic communication,
etc.~\cite{Pikovsky:2002_SynhroBook,Boccaletti:2002_SynchroPhysReport,
Rulkov:1994_PhaseRelation, Pecora:1997_SynchroChaos,
Rznbl:1998_Nature, Rulkov:2002_ChaoticCommunication}.
Traditionally, different types of synchronous behavior of chaotic
oscillators are distinguished. Each of them is characterized by
its own features and may be detected by specific methods which are
different for every synchronous
regime~\cite{Pikovsky:2002_SynhroBook,Boccaletti:2002_SynchroPhysReport}.
The important aim of research is finding the regularities of the
chaotic synchronization regimes and detecting a relationship
between
them~\cite{Boccaletti:2001_UnifingSynchro,%
Brown:2000_ChaosSynchro,Hramov:2004_Chaos}. In particular, we have
shown~\cite{Hramov:2004_Chaos,Aeh:2005_SpectralComponents} that
the different types of the chaotic synchronization behavior of the
flow systems (such as phase synchronization, lag synchronization,
generalized synchronization, complete synchronization) may be
considered as one type of the synchronous dynamics, namely, time
scale synchronization. Obviously, it is important to develop the
further generalization of the chaotic synchronization theory to
detect the common mechanisms resulting in arising the synchronous
behavior.

The aim of this work is to show that the two types of synchronous
behavior of the chaotic oscillators (the generalized
synchronization~\cite{Rulkov:1995_GeneralSynchro,
Pyragas:1996_WeakAndStrongSynchro,Kocarev:1996_GS,%
Rulkov:1996_AuxiliarySystem,Parlitz:1996_PhaseSynchroExperimental,%
Zhigang:2000_GSversusPS,Guan:2005_BistableChaosGS} and the
noise--induced
synchronization~\cite{Fahy:1992_NoiseInfluence,Martian:1994_SynchroNoise,%
Kaulakys:1995_NoiseSynchro,Chen:1996_NoiseSynchro,Khoury:1996_NoiseCircleMaps,%
Ali:1997_NoiseSynchro,Khoury:1998_NoiseSynchro,Shuai:1998_NoiseSynchro,%
Kaulakys:1999_SynchroNoise,Toral:2001_NoiseSynchro,Zhou:PRL2002,Zhou:PRE2003,Zhou:PRE67,Zhou:Chaos2003})
which are traditionally supposed to be different are caused by the
same mechanisms and should be considered as one phenomenon.

The \emph{generalized synchronization} regime (GS) in two
unidirectionally coupled chaotic oscillators means the presence of
a functional relation $\mathbf{u}(t)=\mathbf{F}[\mathbf{x}(t)]$
between the state vectors of the drive $\mathbf{x}(t)$ and the
response $\mathbf{u}(t)$ systems~\cite{Rulkov:1995_GeneralSynchro,
Pyragas:1996_WeakAndStrongSynchro}. This relation may be rather
complicated and the method of detecting it is usually non-trivial.
Depending on the character of this relation $\mathbf{F}[\cdot]$
--- smooth or fractal --- GS is divided into the strong and the weak
ones~\cite{Pyragas:1996_WeakAndStrongSynchro}, respectively. It is
also important to note that the distinct dynamical systems
(including the systems with the different dimension of the phase
space) may be used as the drive and response oscillators.

To detect the generalized synchronization regime the auxiliary
system approach~\cite{Rulkov:1996_AuxiliarySystem} may be used. In
this case the behavior of the auxiliary system $\mathbf{u}(t)$ is
considered together with the response system $\mathbf{v}(t)$ one.
The auxiliary system is equivalent to the response one, but the
initial conditions must be different, i.e.
$\mathbf{v}(t_0)\neq\mathbf{u}(t_0)$, although both
$\mathbf{v}(t_0)$ and $\mathbf{u}(t_0)$ have to belong to the same
basin of chaotic attractors (if there is the multistability in the
system). If GS takes place in the unidirectionally coupled chaotic
oscillators, the system states $\mathbf{u}(t)$ and $\mathbf{v}(t)$
become equivalent after the transient is finished due to the
existence of the relations
$\mathbf{u}(t)=\mathbf{F}[\mathbf{x}(t)]$ and
$\mathbf{v}(t)=\mathbf{F}[\mathbf{x}(t)]$. Thus, the coincidence
of the state vectors of the response and the auxiliary systems
$\mathbf{v}(t)\equiv\mathbf{u}(t)$ is considered as a criterion of
the GS regime presence.

The generalized synchronization regime may also be detected by
means of the conditional Lyapunov exponent
calculation~\cite{Pyragas:1996_WeakAndStrongSynchro}. GS arises in
the system of two unidirectionally coupled chaotic oscillators
only if the highest Lyapunov conditional exponent is
negative~\cite{Pyragas:1996_WeakAndStrongSynchro}.

The \emph{noise--induced
synchronization}~\cite{Fahy:1992_NoiseInfluence,Martian:1994_SynchroNoise,%
Kaulakys:1995_NoiseSynchro,Chen:1996_NoiseSynchro,Kaulakys:1999_SynchroNoise,%
Toral:2001_NoiseSynchro} means that two identical non--coupled
chaotic oscillators $\mathbf{v}(t)$ and $\mathbf{u}(t)$ are driven
by the common external noise $\xi(t)$. The external noise may
result in the consistence of the vector states of the considered
systems after the transient is finished. The noise--induced
synchronization as well as GS may be realized only if all
conditional Lyapunov exponents are
negative~\cite{Pikovsky:1994_Comment_NoiseSynchro,Longa:1996_CLEnegative,%
Zhou:1998_CLEsNoiseSynchro}.

It has been shown in earlier articles that it is not always
possible to observe the noise--induced synchronization in chaotic
oscillators, because in this case the chaotic system must display
particular properties in the phase space (large contraction
region, limited expansion region, and a permanence time that
within the expansion region is greater than in the contraction
region)~\cite{Zhou:PRE2003,Zhou:PRE67}. At the same time it is
necessary to emphasize that biased noise is not a pure
noise--induced transition, and therefore contraction regions in
that case do not play a crucial role.

It is known that there are two similar mechanisms causing
noise--induced synchronization appearance: (i) the external noise
signal $\xi(t)$ has the mean non--zero value that results in
``moving'' the system to the non--chaotic
regime~\cite{Herzel:1995_NoiseSynchroReconsidered,Malescio:1996_NoiseSynchro,%
Gade:1996_NoiseSynchro,Sanchez:1997_NoiseSynchroMeanValue,%
Lorenzo:1999_ColoredNoiseSynchro,PerezMunuzuri:1999_ShuaNoiseSynchro}.
In this case the states of the dynamical systems follow the
external noise $\xi(t)$ in the same way, and, accordingly, they
coincide with each other; (ii) the external noise with the large
amplitude (perhaps, with the zero mean value) moves the image
point corresponding to the system state to the region of the phase
space with the strong dissipation. In other words, the external
noise allows the system to spend more time in the region of the
phase space where the convergence of the phase trajectories takes
place~\cite{Minai:1998_NoiseSynchro,Loreto:1996_NoiseSynchroZeroMean,%
Lai:1998_NoiseSynchroZeroMean,Minai:1999_NoiseSynchro,Rim:2000_NoiseSynchro,Toral:2001_NoiseSynchro,
Zhou:PRL2002,Zhou:PRE2003,Zhou:PRE67,Zhou:Chaos2003}. So, in both
cases the convergence of the phase trajectories and,
correspondingly, the phase flow contraction, play the main role in
the noise--induced synchronization appearance. One can say, that
the noise--induced synchronization is caused by introducing the
additional dissipation into the system either by means of the bias
of the noise or with the help of the large noise amplitude.

The similar effects concerning introducing the additional
dissipation in the system result in the generalized
synchronization regime appearance. As it has been shown in our
works~\cite{Hramov:2005_GSNature,Hramov:CGLE2005}, there are also
two mechanisms causing the GS existence. The first of them is
realized if GS takes place in two systems with unidirectional
dissipative coupling. For such situation the equations describing
the system dynamics may be written as
\begin{equation}
\begin{array} {ll}
\mathbf{\dot x}(t)=\mathbf{H}(\mathbf{x}(t))\\
\mathbf{\dot u}(t)=\mathbf{H}(\mathbf{u}(t))+
\varepsilon\mathbf{A}(\mathbf{x}(t)-\mathbf{u}(t)),
\end{array}
\label{eq:Oscillators1}
\end{equation}
where $\mathbf{A}={\{\delta_{ij}\}}$ is the coupling matrix,
$\varepsilon$ is the control parameter characterizing the coupling
strength between the chaotic oscillators, $\delta_{ii}=0$ or $1$,
$\delta_{ij}=0$ ($i\neq j$). In this case one can see that the
response system $\mathbf{u}(t)$ may be considered as a
\emph{modified system}
\begin{equation}
\mathbf{\dot u}_m(t)=\mathbf{H}'(\mathbf{u}_m(t), \varepsilon)
\label{eq:RsOsc}
\end{equation}
(where $\mathbf{H}'(\mathbf{u}(t))=
\mathbf{H}(\mathbf{u}(t))-\varepsilon\mathbf{A}\mathbf{u}(t)$)
under the external force $\varepsilon\mathbf{A}\mathbf{x}(t)$:
\begin{equation}
\mathbf{\dot u}_m(t)=\mathbf{H}'(\mathbf{u}_m(t),\varepsilon)+
\varepsilon\mathbf{A}\mathbf{x}(t), \label{eq:RsOsc&Force}
\end{equation}
It is easy to see that the term
$-\varepsilon\mathbf{A}\mathbf{u}(t)$ brings the additional
dissipation into the system~(\ref{eq:RsOsc}). Indeed, the phase
flow contraction is characterized by means of the vector field
divergence. Obviously, the vector field divergences of the
modified and the response systems are related with each other as
\begin{equation}
\mathrm{div}\,\mathbf{H}'=\mathrm{div}\,\mathbf{H}
-\varepsilon\sum\limits_{i=1}^N\delta_{ii}
\end{equation}
(where $N$ is the dimension of the modified system phase space),
respectively. So, the dissipation in the modified system is
greater than in the response one and it increases with growth of
the coupling strength $\varepsilon$.

The generalized synchronization regime arising
in~(\ref{eq:Oscillators1}) may be considered as a result of two
cooperative processes taking place simultaneously. The first of
them is the growth of the dissipation in the
system~(\ref{eq:RsOsc}) and the second one is an increase of the
amplitude of the external signal. Evidently, both processes are
correlated with each other by means of parameter $\varepsilon$ and
can not be realized in the coupled oscillator
system~(\ref{eq:Oscillators1}) independently. Nevertheless, it is
clear, that an increase of the dissipation in the modified
system~(\ref{eq:RsOsc}) results in the simplification of its
behavior and the transition from the chaotic oscillations to the
periodic ones. Moreover, if the additional dissipation is large
enough the stationary fixed state may be realized in the modified
system. On the contrary, the external chaotic force
$\varepsilon\mathbf{A}\mathbf{x}(t)$ tends to complicate the
behavior of the modified system and impose its own dynamics on it.
Obviously, the generalized synchronization regime may not appear
unless own chaotic dynamics of the modified system is suppressed.

One can see that in this case the reasons resulting in the
generalized synchronization arising are very similar to the
mechanisms which may be revealed for the noise-induced
synchronization with biased noise. Indeed, as well as in the case
of the biased noise the system state is moved by the deterministic
effect to the non-chaotic regime and, as result, the generalized
regime may be detected.

The second mechanism of GS arising is realized when two
oscillators are coupled in the unidirectional non--dissipative
way. In this case the signal of the master oscillator should be
introduced with the large amplitude into the response system. This
signal moves the response system state in the region of the phase
space with the strong dissipation (see,
e.g.~\cite{Hramov:2005_GSNature}) as well as in the case of the
noise--induced synchronization. Both mechanisms of GS arising are
characterized by the convergence of the phase trajectories and all
conditional Lyapunov exponents are negative in these cases. It
should be noted that in~\cite{Hramov:CGLE2005} it was shown that
both mechanisms lead to the GS regime onset simultaneously.

So, one can see, that the noise--induced synchronization and the
generalized synchronization regimes are caused by the same
mechanism. In most cases this mechanism is the suppression of own
chaotic dynamics of the response system by means of the non--zero
mean of the noise, or with the help of the additional dissipative
term, or by moving the system state into the regions of the phase
space with the strong convergence of the phase trajectories. It
should be noted, that it is not a rigorous mathematical proof, but
the given arguments seem to be quite convincing for understanding
the unified character of these two phenomena.

The equivalence of these two types of the synchronous behavior may
also be illustrated by the following conclusion: the
noise--induced synchronization regime means the presence of the
functional relationship $\mathbf{F}[\cdot]$ between the chaotic
oscillator state and the stochastic signal. Indeed, two identical
systems $\mathbf{u}(t)$ and $\mathbf{v}(t)$ driven by the common
stochastic force $\xi(t)$ in the regime of the noise--induced
synchronization behave equivalently, i.e.,
${\mathbf{u}(t)=\mathbf{v}(t)}$. Obviously,
${\mathbf{u}(t)=\mathbf{F}_u[\xi(t)]}$ and
${\mathbf{v}(t)=\mathbf{F}_v[\xi(t)]}$, where
$\mathbf{F}_u[\cdot]$ and $\mathbf{F}_v[\cdot]$ are some
functional dependences, distinct for the different initial
conditions. Nevertheless, in the noise--induced synchronization
regime after the transient is finished the vector states of
considered systems coincide with each other, therefore,
$\mathbf{F}_u[\cdot]\equiv\mathbf{F}_v[\cdot]\equiv\mathbf{F}[\cdot]$
independently on the initial conditions. So, in the case of the
noise--induced synchronization the following functional relation
takes place: $\mathbf{u}(t)=\mathbf{v}(t)=\mathbf{F}[\xi(t)]$. The
same statement is used for the generalized synchronization
definition, when the response system is driven by the chaotic
signal instead of the stochastic one.

Let us show, that the generalized synchronization regime may be
obtained if the drive chaotic system is replaced by the noise
signal. This effect may also be treated as the noise--induced
synchronization. As the first example of such system behavior let
us consider the unidirectionally coupled logistic maps
\begin{equation}
\begin{array} {ll}
& x_{n+1}=f(x_n), \\
& y_{n+1}=f(y_n)+\varepsilon(f(x_n)-f(y_n)),
\end{array} \label{eq:LogMaps}
\end{equation}
where $f(x)=ax(1-x)$, $a$ is the control parameter, $\varepsilon$
--- the coupling strength. The presence of the GS regime in this system for some values of
the coupling strength $\varepsilon$ has been shown
(see~\cite{Pyragas:1996_WeakAndStrongSynchro}). Let us consider
now the behavior of the response system $y_n$ when the dynamics of
$x$ variable is not determined by the dynamical
system~(\ref{eq:LogMaps}), but it is the stochastic process
$\xi_n$ which is characterized by the probability distribution
$p(\xi)$. In this case the dynamics of the response system is
described by the equation
\begin{equation}
y_{n+1}=f(y_n)+\varepsilon(f(\xi_n)-f(y_n)).
\label{eq:StochLogMap}
\end{equation}
We have shown that the synchronous dynamics between the stochastic
process and the state of the dynamical system can also take place
in spite of the random character of $\xi$ as well as in the cases
of the generalized synchronization or the synchronization induced
by the noise. This effect is very similar to the noise induced
synchronization with biased noise although the movement of the
system state into non-chaotic regime is caused by the term
$-\varepsilon f(y_n)$ instead of the bias of noise.

\begin{figure}[tb]
\vspace*{0.3cm}
\centerline{\includegraphics*[scale=0.5]{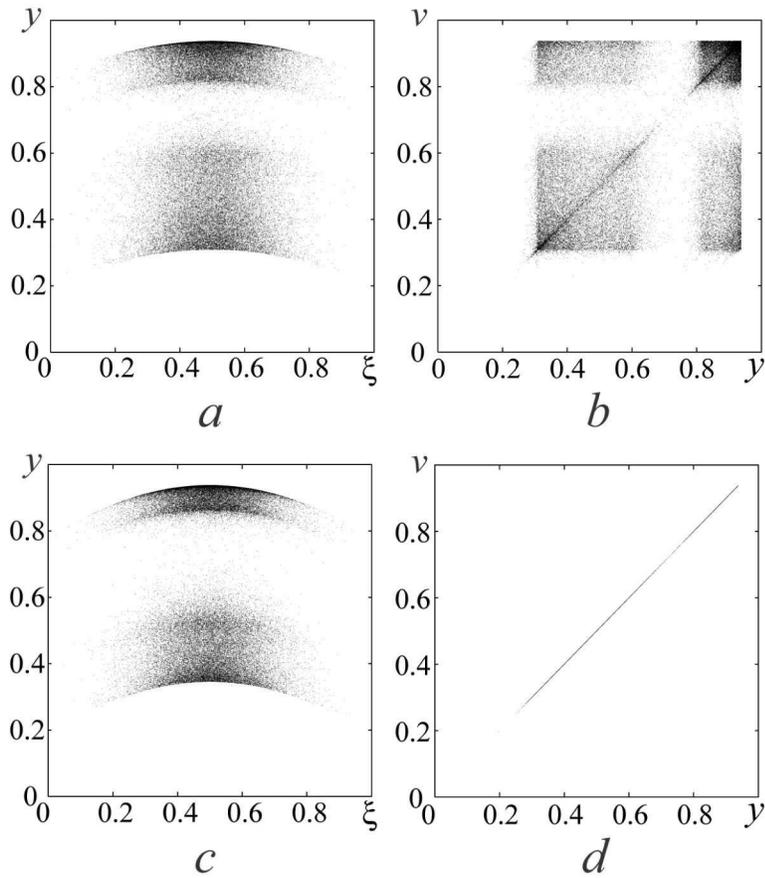}} \caption{The
planes $(\xi_n,y_n)$ and $(y_n,v_n)$ of the logistic maps for the
coupling strength $\varepsilon=0.125$ (\textit{a, b}) and
$\varepsilon=0.175$ (\textit{c, d}). It is clear that in the case
(\textit{d}) the response $y_n$ and the auxiliary $v_n$ systems
demonstrate the identical behavior $y_n=v_n$ that testifies the
presence of the functional relationship ${y_n=\mathbf{F}[\xi_n]}$,
and, therefore, the establishment of the synchronization regime}
\end{figure}

To detect the presence of the relationship between the stochastic
process $\xi_n$ and the state $y_n$ of the dynamical system we
have used the auxiliary system approach described above. The
behavior of the response and the auxiliary systems is shown in
Fig.~1,\,\textit{b} when the parameters have been selected as
$a=3.75$ and $\varepsilon=0.125$, the probability distribution of
the random variable $\xi_n$ is
\begin{equation}
p(\xi)=\frac{1}{\sqrt{2\pi}\sigma}\exp\left(-\frac{(\xi-\xi_0)^2}{2\sigma^2}\right),
\label{eq:NormalDistrib}
\end{equation}
where $\xi_0=1/2$, $\sigma=0.11$\footnote{It is important to note
that the character of the distribution of the random variable
$\xi$ does not matter and the similar results may be observed for
the others types of the probability distribution $p(\xi)$, for
example, for the uniform one.}.

It is clear, the response and the auxiliary systems are
characterized by the different states in the same moment of
discrete time when the coupling strength is small enough
($\varepsilon=0.125$). The points corresponding to the states of
the response and the auxiliary systems are spread over all area
$(y_n, v_n)$. It means that there are no functional relation
between the stochastic process $\xi_n$ and the state $y_n$ of the
dynamical system.

With increasing the coupling strength ($\varepsilon=0.175$) the
behavior of the considered system is radically changed (see
Fig.~1,\,\textit{d}). The points corresponding to the state of the
considered systems are on the straight line $v_n=y_n$. Therefore,
the relationship $y_n=\mathbf{F}[\xi_n]$ takes place and the
synchronous behavior is observed. It is important to note, that
the functional relationship $\mathbf{F}[\cdot]$ is fractal (see
Fig.~1,\,\textit{c}) that corresponds to the case of the weak
synchronization~\cite{Pyragas:1996_WeakAndStrongSynchro}. Nobody
can detect the presence of the functional relationship between
$\xi_n$ and $y_n$ taking into account $(\xi,y)$--plane only
(compare Fig.~1,\,\textit{a} when the synchronous regime is not
observed and and Fig.~1,\,\textit{c} when the synchronization
takes place, respectively).

\begin{figure}[tb]
\vspace*{0.3cm}
\centerline{\includegraphics*[scale=0.5]{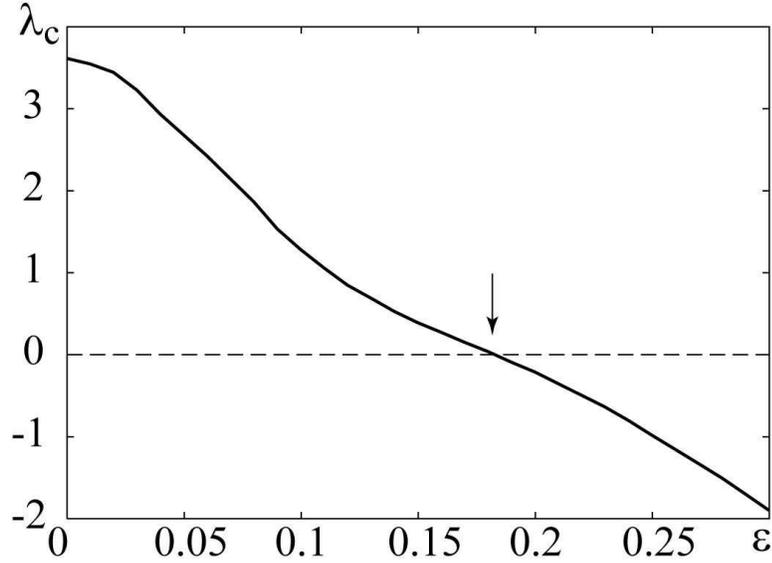}} \caption{The
dependence of the conditional Lyapunov exponent $\lambda_c$ of the
system~(\ref{eq:StochLogMap}) on the coupling strength
$\varepsilon$. The stochastic signal is characterized by the
normal distribution (\ref{eq:NormalDistrib}), the onset of the
synchronization is marked by an arrow}
\end{figure}

The presence of the synchronous regime is also confirmed by the
dependence of the conditional Lyapunov exponent $\lambda_c$ on the
coupling strength $\varepsilon$ (Fig.~2). One can see that
$\lambda_c$ is positive for the small values of the coupling
parameter, therefore there is no the functional relationship
between $\xi_n$ and $y_n$. When the coupling strength increases
the conditional Lyapunov exponent $\lambda_c$ becomes negative,
therefore, the synchronous regime is detected and the relationship
$y_n=\mathbf{F}[\xi_n]$ between stochastic process $\xi_n$ and the
state $y_n$ of the logistic map~(\ref{eq:StochLogMap}) takes
place.

The analogous results have been obtained for the R\"ossler system
under the external stochastic signal. As in the case of the first
example~(\ref{eq:StochLogMap}) let us replace the dynamics of the
drive system by the stochastic process $\xi_n$. The investigated
system has the following form:
\begin{equation}
\begin{array} {ll}
& \dot x_r = -\omega_r y_r-z_r+\varepsilon(\xi_n-x_r), \\
& \dot y_r = \omega_r x_r+a y_r, \\
& \dot z_r = p+z_r(x_r-c),
\end{array} \label{eq:Roesslers}
\end{equation}
where $a=0.15$, $p=0.2$, $c=10.0$, $\omega_r=0.95$ are the control
parameter values,  $\xi_0=0$, $\sigma=11.2$ --- mean value and
dispersion of probability distribution
function~(\ref{eq:NormalDistrib}) of random value $\xi_n$,
respectively. As well as for the logistic map with noise term (see
equation~(\ref{eq:StochLogMap})) the observed effect is similar to
the noise-induced synchronization with biased noise.

For $\varepsilon=0.05$ (see Fig.~3, \textit{a, b}) the
noise--induced synchronization does not observed, i.e. all points
on $(x,v)$--plane characterizing response and auxiliary systems
states are spread randomly. When the coupling parameter increases
($\varepsilon=0.15$), the response and auxiliary systems
demonstrate identical behavior (see Fig.~3, \textit{d}). This
situation also corresponds to the case of weak GS synchronization
(see Fig.~3, \textit{c}). It should be noted that the external
stochastic signal has been introduced in
system~(\ref{eq:Roesslers}) in the way that is typical for the
mutually coupled oscillators when the GS regime takes place.
Alternatively, this coupling term is not typical for the system
where noise--induced synchronization is observed. We think that
this example is an additional argument confirming our conclusion.

\begin{figure}[tb]
\vspace*{0.3cm}
\centerline{\includegraphics*[scale=0.5]{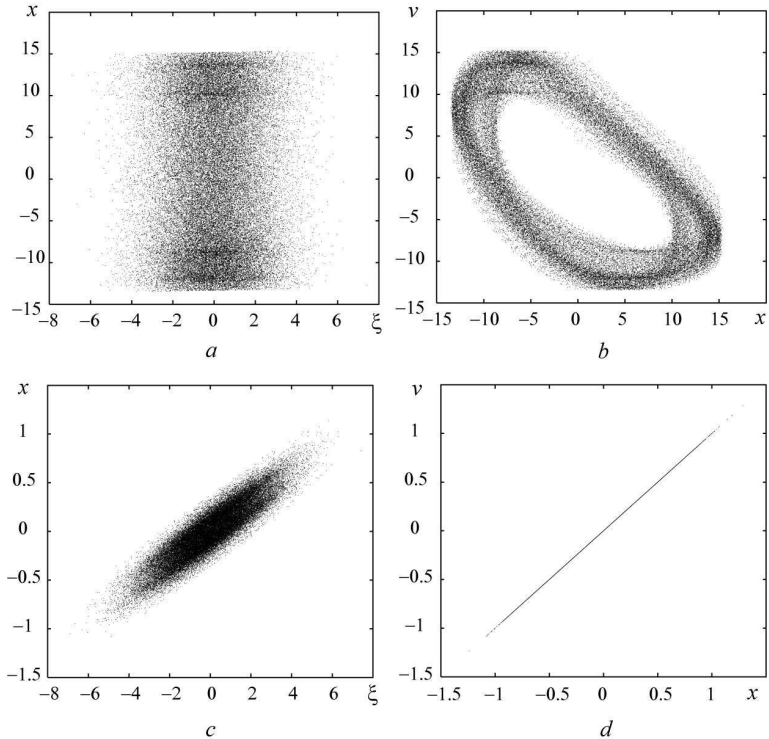}} \caption{The
planes $(\xi_n,x)$ and $(x,v)$ of the R\"ossler systems for the
coupling strength $\varepsilon=0.05$ (\textit{a, b}) and
$\varepsilon=0.15$ (\textit{c, d}). It is clear that in the case
(\textit{d}) the synchronization regime is observed}
\end{figure}

In conclusion, we argue that the generalized synchronization and
the noise--induced synchronization regimes are caused by the same
mechanism. As it has been mentioned above, this mechanism is the
suppression of own chaotic dynamics of the response system by
means of introducing the additional dissipation. The additional
dissipation may be introduced into the system either by means of
the mean non--zero value of the noise, or with the help of the
additional dissipative term, or by moving the system state into
the regions of the phase space with the strong convergence of the
phase trajectories. Typically, the mechanisms resulting in the
generalized synchronization act like ones in the case of the
noise-induced synchronization with biased noise when the system
state is moved (by means of the dissipative term or biased noise)
to the non-chaotic regime. Nevertheless, the other mechanism
corresponding to the movement of the system state into the regions
of the phase space with the strong dissipation by means of the
external signal with large amplitude or by means of large
zero-mean noise may also take place (see,
e.g.~\cite{Toral:2001_NoiseSynchro,Pyragas:1996_WeakAndStrongSynchro}).

So, the difference between the generalized synchronization and the
noise--induced synchronization is only in character of the driving
signal. In case of the noise--induced synchronization the
stochastic signal drives the chaotic oscillator, while in the case
of the generalized synchronization the signal of another chaotic
dynamical system is used. That is why the system with the
different dimensions of the phase space may be used to obtain the
generalized synchronization regime. Obviously, the identity of the
system is not required in this case and, in general, the driving
signal may be arbitrary. Although the generalized synchronization
and the noise--induced synchronization are traditionally
distinguished as different types of the synchronous behavior, it
may be appropriate and useful to consider them as one type of the
synchronous behavior caused by one reason.


We thank Dr. Svetlana V. Eremina for the English language support.
This work has been supported by U.S.~Civilian Research \&
Development Foundation for the Independent States of the Former
Soviet Union (CRDF), Grant No. {REC--006}), Russian Foundation of
Basic Research (Projects 05--02--16273, 06--02--16451) and the
Program of President of Russion Federation for the Support of the
Leading Scientific Schools (NSh-4167.2006.2). We also thank the
``Dynastiya'' Foundation for the financial support.




\newpage
\bibliographystyle{elsart-num}
\bibliography{Thesis}




\end{document}